\documentclass[aip,jcp,superscriptaddress,amsfonts,graphicx,preprint,numerical]{revtex4-2}
\usepackage[utf8]{inputenc}
\usepackage[T1]{fontenc}
\usepackage{tgtermes}
\usepackage{amsmath, amssymb}
\usepackage{mathtools}
\usepackage{breqn}
\usepackage{graphicx}
\usepackage{float}
\usepackage{bm}
\usepackage{physics}
\usepackage{xcolor}
\usepackage{verbatim}
\usepackage{enumitem}

\usepackage{hyperref}
\usepackage{multirow,xcolor}
\hypersetup{
    colorlinks = true,
    linkcolor = blue,
    linkbordercolor = white,
    citecolor=blue,
    urlcolor=black
}
\usepackage[export]{adjustbox}
\usepackage{subfig,caption}
\newcommand{\red}[1]{#1}

\newcommand{\Rb}{\mathbf{R}}
\newcommand{\Pb}{\mathbf{P}}

\makeatletter
\let\cat@comma@active\@empty
\makeatother

\begin{document}
\title{Linear and Angular Momentum Conservation in Surface Hopping Methods}
\author{Yanze Wu}
\email{wuyanze@sas.upenn.edu}
\affiliation{Department of Chemistry, University of Pennsylvania, Philadelphia, Pennsylvania 19104, USA}
\author{Jonathan Rawlinson}
\affiliation{Department of Mathematics, University of Manchester, Manchester M13 9PL, UK}
\author{Robert G. Littlejohn}
\affiliation{Department of Physics, University of California, Berkeley, California 94720, USA}
\author{Joseph E. Subotnik}
\email{subotnik@sas.upenn.edu}
\affiliation{Department of Chemistry, University of Pennsylvania, Philadelphia, Pennsylvania 19104, USA}
\date{\today}

\begin{abstract}
    We demonstrate that, for systems with spin-orbit coupling and an odd number of electrons, the standard fewest switches surface hopping (FSSH) algorithm does not conserve the total linear or angular momentum. This lack of conservation arises not so much from the hopping direction (which is easily adjusted) but more generally from propagating adiabatic dynamics along surfaces that are not time reversible. We show that one solution to this problem is to run along eigenvalues of phase-space electronic Hamiltonians $H(\Rb,\Pb)$ (i.e. electronic Hamiltonians that depend on both nuclear position and momentum) with an electronic-nuclear coupling $\bm{\Gamma} \cdot \mathbf{P}$ (see Eq.~\eqref{eq:HnaGamma}) and we delineate the conditions that must be satisfied by the operator $\bm{\Gamma}$. The present results should be extremely useful as far as developing new semiclassical approaches that can treat systems where the nuclear, electronic orbital, and electronic spin degrees of freedom altogether are all coupled together, hopefully including systems displaying the chiral induced spin selectivity (CISS) effect.
\end{abstract}

\maketitle

\section{Introduction}

Nonadiabatic processes that violate the Born-Oppenheimer approximation are prevalent in physical and chemical dynamics, including photochemical and charge transfer reactions. Quite often, the electronic spin is an important ingredient (and not an innocent bystander) that can facilitate an important relaxation process: intersystem crossing (ISC)\cite{tretiak:2013:scirep_isc,prezhdo:2012:isc_nanotubes,marian:2012:wires_isc}. Now, while many chemists routinely discuss triplet versus singlet dynamics (distinguishing spin state by their total spin [$S^2$] eigenvalue), it is worth noting that in systems with reasonably strong spin-orbit coupling or in a strong magnetic field \cite{pemberton2022Revealing}, the spin direction ($m_s$) can also be quite important; for instance, $S_z$ can play an important role in maintaining total angular momentum conservation, $\mathbf{L}_{\text{molecule}} = \mathbf{L}_{\text{nuclear}} + \mathbf{L}_{\text{electron}} + \mathbf{S}_{\text{electron}}$. In fact, recent {\em ab initio} studies \cite{bistoni2021Intrinsic,tao2023Symmetric} have pointed out that spin cannot be ignored when running molecular dynamics if one wishes to conserve angular momentum, and in principle spin-dependent nuclear motion is measurable and can have strong consequences \cite{bistoni2021Intrinsic,tao2023Symmetric,wu2021Electronic,fransson2023Chiral}.

When simulating nonadiabatic dynamics, one must inevitably make approximations on account of computational cost. To that end, Tully's fewest switch surface hopping (FSSH) \cite{tully1990Molecular} is perhaps the most widely used approach in practice. Since the framework of surface hopping does not depend on the specific electronic Hamiltonian, one might presume that FSSH can directly model systems with different $m_s$ quantum numbers (e.g. triplet states) simply by expanding the electronic Hilbert space to include all three basis functions in the triplet subspace; indeed, several research groups \cite{mai2018Nonadiabatic,gonzalez:2014:pccp:uracil_isc,thiel:2014:jcp:fssh_isc,tavernelli:2015:jcp:fssh_tddft_isc,prezhdo:2012:isc_nanotubes} have successfully run such dynamics to look at photochemical problems. We will show below, however, that these studies do not conserve the total (nuclear plus electronic) angular momentum either during propagation or during a hop (in agreement with a recent finding by Shu {\em et al} \cite{shu2020Conservation} who investigated FSSH without SOC). In general, on account of this finding, it is clear that one must be cautious when analyzing the details of which $m_s$ spin state relaxes in which way. More generally, without a proper treatment of conservation laws, there is no way to confidently apply the surface hopping approach to study interesting physics at the intersection of spintronics and dynamics, e.g. the chirality induced spin selectivity (CISS) effects \cite{naaman2019Chiral} or the dynamics of spin-dependent chiral phonons \cite{zhang2015Chiral,fransson2023Chiral}.

In the present paper, we will directly address this lack of momentum conservation  in the context of the FSSH algorithm and we will isolate the underlying problem. 
Most importantly, we will show below that FSSH can be fixed up to conserve linear and angular momentum by building an electronic Hamiltonian (to be diagonalized) that depends on the nuclear velocity, leading to so-called phase-space surface hopping (PSSH). For the seasoned reader, in Eqs.~\eqref{eq:GammaTcov}-\eqref{eq:GammagradRcov} below, we show the necessary conditions required for PSSH methods to conserve momentum. We believe that the present manuscript should pave the way for new hopping algorithms that conserve momentum and automatically incorporate the molecular Berry curvature effects \cite{mead1979determination,berry1993Chaotic,bian2021Modeling,monzel2022Molecular,li2022Energy}. 

An outline of this paper is as follows. In Sec.~\ref{sec:symmetry1}, we provide the reader with the relevant background needed: we present the fine-structure Hamiltonian, we define the relevant definitions of momentum/angular momentum operators in the context of mixed quantum-classical frameworks, and we discuss how the relevant matrix elements behave under translations and rotations. In Sec.~\ref{sec:fssh}, we reconsider the standard surface hopping algorithm, and demonstrate conclusively that the algorithm does not satisfy either linear or angular momentum conservation. The heart of this manuscript is Sec.~\ref{sec:pssh}, where we show that certain phase-space generalizations of FSSH (PSSH) can in fact recover linear and angular momentum conservation, and we explicitly list the conditions that must be satisfied in order to maintain such conservation. We further discuss the nuances of hopping directions within a PSSH scheme. In Sec.~\ref{sec:conclusion}, we conclude and point out some key future directions for this research. Notably, in Appendix \ref{sec:pssh_symdrvgrad}, we connect the main body of the text presented here with the original PSSH algorithm proposed by Shenvi \cite{shenvi2009Phasespace}.

Before concluding, given the many different degrees of freedom inherent in a mixed quantum-classical algorithm, we list below (in Table.~\ref{table:symbol}) our indices and nomenclature:
\begin{table}[H]
\begin{center}
\begin{tabular}{c|l}
\hline
Symbol & Denotes \\ \hline
$\alpha,\beta,\gamma,\delta,\zeta$ & Spatial directions ($x,y,z$) \\
$I,J$ & Nuclear indices \\
$a,b$ & Electronic indices \\
$j,k$ & Adiabats \\
$\tilde{k}$ & The active adiabat in FSSH \\
$m,n$ & Phase-space adiabats \\
$\tilde{n}$ & The active phase-space adiabat in PSSH \\
\hline
\end{tabular} \caption{List of Indices}\label{table:symbol}
\end{center}
\end{table}
\noindent Vectors in 3 or 3N dimensional space are written in boldface.

\section{Background} \label{sec:symmetry1}

Below, we will work with the standard molecular Born-Oppenheimer (BO) Hamiltonian that includes electrostatic interactions and spin-orbit coupling:
\begin{align}
    \hat{V} = \sum_{I,J}{\frac{Q_IQ_J}{\abs{\mathbf{R}_I-\mathbf{R}_J}}} - \sum_{I,a}{\frac{Q_I}{\abs{\mathbf{R}_I-\hat{\mathbf{r}}_a}}} + \sum_{a,b}{\frac{1}{\abs{\hat{\mathbf{r}}_a-\hat{\mathbf{r}}_b}}} -\sum_a{\frac{\hat{\mathbf{p}}_a\cdot\hat{\mathbf{p}}_a }{2m_e}} + \hat{V}_{SO} \label{eq:V}
\end{align}
where the SOC term is
\begin{align}
    \hat{V}_{SO} = \frac{Q_I}{c^2}\sum_{I,a}{\left(\frac{\hat{\mathbf{r}}_a-\mathbf{R}_I}{\abs{\hat{\mathbf{r}}_a-\mathbf{R}_I}^3}\times\hat{\mathbf{p}}_a \right)\cdot\mathbf{\hat{s}}_a} \label{eq:Vso}
\end{align}
Here, the $Q_I,Q_J$ are the nuclear charges, \red{$\mathbf{R}$ is the nuclear coordinate, and $\hat{\mathbf{r}},\hat{\mathbf{p}}$ and $\hat{\mathbf{s}}$ are the electronic position, momentum and spin operators, respectively. Throughout this paper, we will use the hat notation ($\hat{\phantom{a}}$) to represent electronic operators.}

Note that, in Eqs.~\eqref{eq:V} and \eqref{eq:Vso}, we have summed over all electrons. Below and henceforward, it will be convenient to switch to a second-quantized formalism where we replace, e.g., the momentum \red{operator} for a single electron with the momentum operator for all of the electrons (in Fock space):
\begin{equation}
\begin{split}
\hat{p}_{\alpha} &\equiv \sum_{a}{\hat{p}_{a,\alpha}} \\ \hat{l}_{\alpha} &\equiv \sum_{a}{\hat{l}_{a,\alpha}} \\ \hat{s}_{\alpha} &\equiv \sum_{a}{\hat{s}_{a,\alpha}}   
\end{split}    
\end{equation}
\red{Here, $a$ indexes the individual electrons.} 

Before we address momentum conservation, we will now review several definitions and symmetry properties as relevant for a quantum mechanical system.

\subsection{Mixed Quantum-Classical Definition of Linear and Angular Momentum}
Within a mixed quantum-classical framework, the total {\em nuclear} linear momentum and angular momentum are defined by
\begin{align}
    P_{nu,\alpha} &= \sum_I{M_I\dot{R}_{I\alpha}} \label{eq:Pnu} \\
    L_{nu,\alpha} &= \sum_{I,\beta,\gamma}{\epsilon_{\alpha\beta\gamma}M_IR_{I\beta}\dot{R}_{I\gamma}} \label{eq:Lnu}
\end{align}
where $R_{I\alpha}$ is the coordinate of atom $I$ in direction $\alpha$, $M_I$ is the nuclear mass, and $\epsilon_{\alpha\beta\gamma}$ is the Levi-Civita symbol. The total {\em molecular} linear momentum and angular momentum are defined by summing over the nuclear  and electronic quantities:
\begin{align}
    P_{mol,\alpha} &= \sum_I{M_I\dot{R}_{I\alpha}} + \mel{\psi}{\hat{p}_\alpha}{\psi} \label{eq:Pmol} \\
    L_{mol,\alpha} &= \sum_{I,\beta,\gamma}{\epsilon_{\alpha\beta\gamma}M_IR_{I\beta}\dot{R}_{I\gamma}} + \mel{\psi}{\hat{l}_\alpha + \hat{s}_\alpha)}{\psi} \label{eq:Lmol}
\end{align}
where $\psi$ is the electronic wavefunction, and $\hat{p}$, $\hat{l}$ and $\hat{s}$ are the single-body electronic momentum, orbital angular momentum and electronic spin operators, respectively. 

\subsection{The Behavior of the Hamiltonian and the Set of Adiabatic States Under Translations and Rotations}

In this paper, we restrict ourselves to finite systems which can be translated and rotated in free space without any change in the fundamental physics. To that end, note that the BO Hamiltonian $\hat{V}$ in Eq.~\eqref{eq:V} is invariant to the {\em total} (nuclear + electronic) translation/rotation of the system; one cannot translate or rotate the individual nuclear/electronic components without changing the physics. To explore the consequences of these symmetries, let us define total nuclear momentum and angular momentum operators
\begin{align}
    \mathcal{P}_\alpha &= -i\hbar \sum_I{\pdv{}{R_{I\alpha}}} \label{eq:Pop} \\
    \mathcal{L}_\alpha &= -i\hbar \sum_{I,\beta,\gamma}{\epsilon_{\alpha\beta\gamma}R_{I\beta}\pdv{}{R_{I\gamma}}} \label{eq:Lop}
\end{align}
By translational/rotational invariance,  the Hamiltonian presented in Eq.~\eqref{eq:V} satisfies
\begin{align}
    [\hat{V},\mathcal{P}_\alpha + \hat{p}_\alpha] &= 0 \label{eq:VTinv} \\
    [\hat{V},\mathcal{L}_\alpha + \hat{l}_\alpha + \hat{s}_\alpha] &= 0
\end{align}

Now, the BO picture defines a set of adiabatic basis $\ket{k(\mathbf{R})}$, which are electronic wavefunctions parameterized by the nuclear coordinates $\mathbf{R}$. Importantly, the BO framework is not compete without defining the phases of the relevant adiabatic states. As pointed out by Littlejohn \cite{littlejohn2023Representation}, momentum conservation makes the most sense if one chooses the adiabat states $\ket{k}$ to have phases defined as follows:
\begin{align}
    (\mathcal{P}_\alpha + \hat{p}_\alpha)\ket{k}&=0 \label{eq:Tcov} \\
    (\mathcal{L}_\alpha + \hat{l}_\alpha + \hat{s}_\alpha)\ket{k}&=0 \label{eq:Rcov}
\end{align}
for all $\alpha$. 
Eq.~\eqref{eq:Tcov} and \eqref{eq:Rcov} dictate that one chooses well-defined phases of the adiabatic states that are functions of only the relative (not absolute) coordinates of the nuclei and electrons. As we discuss in Appendix \ref{sec:adiabat_cov} (and as shown in Ref. \cite{littlejohn2023Representation}), Eqs.~\eqref{eq:Tcov} and \eqref{eq:Rcov} are always valid provided one applies a bra $\bra{j} (j \ne k)$ to these equations; however, a phase convention is necessary if we want these equations to hold in general (with $j = k$).

\subsection{The Behavior of the Hamiltonian Gradients and Derivative Couplings Under Various Symmetries}

In this section, we will derive a few symmetry properties of the Hamiltonian gradients $\nabla V_{jk} = \nabla\mel{j}{\hat{V}}{k}$ and derivative couplings $\mathbf{d}_{jk} = \mel{j}{\nabla}{k}$, when the phases of the basis states $\ket{j}$ and $\ket{k}$ follow the phases in Eqs.~\eqref{eq:Tcov} and \eqref{eq:Rcov} for translational and rotational motion.
We begin with the derivative couplings. By projecting Eqs.~\eqref{eq:Tcov} and \eqref{eq:Rcov} to some other state $\bra{j}$, we find that $\mathbf{d}_{jk}$ satisfies
\begin{align}
    -i\hbar\sum_I{d_{jk}^{I\alpha}} + \mel{j}{\hat{p}_\alpha}{k} &= 0 \label{eq:dcTcov} \\
    -i\hbar\sum_{I,\beta,\gamma}{\epsilon_{\alpha\beta\gamma}R_{I\beta}d_{jk}^{I\gamma}} + \mel{j}{\hat{l}_\alpha + \hat{s}_\alpha}{k} &=0 \label{eq:dcRcov}
\end{align}
which indicates that the derivative coupling between two states has both a translational and a rotational component. Summed over nuclei, these translational and rotational components are equal to the transition electronic momentum\cite{fatehi2012Derivative} and angular momentum matrix elements\cite{yarkony:1989:jcp_emailme_ang}.

\red{As for the gradients, in Appendix \ref{sec:gradinvproof}, we show that these matrix elements satisfy:}
\begin{align}
    \sum_{I}{\nabla_{I\alpha}V_{jk}} = 0 \label{eq:bogradTinv} \\
    \sum_{I,\beta,\gamma}{\epsilon_{\alpha\beta\gamma}R_{I\beta}\nabla_{I\gamma}V_{jk}} = 0 \label{eq:bogradRinv}
\end{align}
\red{Eqs.~\eqref{eq:bogradTinv}-\eqref{eq:bogradRinv} will be helpful for proving the relevant conservation laws for various surface hopping methods below.}

\section{Standard FSSH and Momentum Conservation} \label{sec:fssh}

Let us now briefly review the algorithm of the standard FSSH. FSSH spawns a swarm of trajectories, each associated with a nuclear coordinate $\mathbf{R}$, a nuclear momentum $\mathbf{P}$, an active adiabatic surface $\tilde{k}$ and an electronic amplitude $c_k$ on adiabat $k$. At each timestep, these quantities are propagated by (here we assume the EOMs are written in Cartesian coordinates)
\begin{align}
    \dot{R}_{I\alpha} &= \frac{P_{I\alpha}}{M_I} \\
    \dot{P}_{I\alpha} &= -\ev{\nabla_{I\alpha} \hat{V}}{\tilde{k}} \\
    \dot{c}_j &= -\frac{i}{\hbar}E_jc_j - \sum_{I,\alpha,k}{\dot{R}_{I\alpha} d_{jk}^{I\alpha} c_k}
\end{align}
At each timestep, the trajectory has a chance to change its active surface (``hop''). The hopping rate from surface $j\to k$ is given by $g_{j\to k} = \max(2\Re[\sum_{I,\alpha}{\dot{R}_{I\alpha} d^{I\alpha}_{jk}}c_k/c_j],0)$.
As derived by Pechukas \cite{pechukas1969Timedependent}, Herman \cite{herman1984Nonadiabatic}, Kapral \cite{kapral2016Surface}, and Tully \cite{tully1991Nonadiabatic}, at each successful hop from $j\to k$, the momentum is rescaled along the direction of the derivative coupling $\mathbf{d}_{jk}$ to conserve energy. 

At this point, we have enough background to prove that a naive implementation of FSSH does not conserve either the total linear momentum $\mathbf{P}_{mol}$ or the total angular momentum $\mathbf{L}_{mol}$. 

\subsection{Linear and Angular Momentum During Motion Along A Single Surface}
When running without a hop, FSSH is equivalent to Born-Oppenheimer dynamics, where only the total nuclear quantities $\mathbf{P}_{nu}$ and $\mathbf{L}_{nu}$ are conserved (Ref.~\cite{littlejohn2023Representation}). In such a case, the molecular quantities $\mathbf{P}_{mol}$ or $\mathbf{L}_{mol}$ defined in Eqs.~\eqref{eq:Pmol} and \eqref{eq:Lmol} will be conserved only when the electrons have vanishing expectation values of momentum $\left( \ev{\hat{\mathbf{p}}}{\tilde{k}}\right)$ or angular momentum $\left(\ev{\hat{\mathbf{l}}+\hat{\mathbf{s}}}{\tilde{k}}\right)$.

More generally, however, there is no reason to assume that these expectation values need to be zero. In particular, non-vanishing expectation values will arise when the surface of interest lacks of time reversibility, e.g., a degenerate surface corresponding to a system with an odd number of electrons. In such a case, it is well known that $\ev{\hat{\mathbf{s}}}\ne 0$ and so classical BO dynamics will not conserve the total angular momentum. As a side note, in a recent article\cite{tao2023Symmetric}, we showed that, in order to maintain angular momentum conservation, one possible approach is to include the Berry force (i.e. the pseudo-magnetic force arising from the Berry curvature), $f^{Berry}_{I\alpha} = \sum_{J,\beta}{(\nabla_{I\alpha} d^{J\beta}_{kk} - \nabla_{J\beta} d^{I\alpha}_{kk})\dot{R}_{J\beta}}$. However, as we will show below, there is a more natural approach to achieve angular momentum conservation that is easier to include within a surface hopping formalism (that does not require an arbitrary choice of any doublet).

\subsection{Linear and Angular Momentum During a Hop}

During the course of a hop in FSSH, the nuclear momentum is rescaled along the direction of the derivative coupling (here we assume a hop from $j\to k$):
\begin{align}
    \mathbf{P}\to\mathbf{P}+\hbar\eta\mathbf{d}_{jk} \label{eq:rescale}
\end{align}
where $\eta$ is a real valued (one-dimensional) amplitude that must be calculated on the fly.

The rescaling in Eq.~\eqref{eq:rescale} can easily violate the relevant conservation laws. In particular, since the naively calculated derivative couplings generally satisfy Eqs.~\eqref{eq:dcTcov} and \eqref{eq:dcRcov}, they necessarily have some translational and rotational component. Therefore, neither $\mathbf{P}_{mol}$ nor $\mathbf{L}_{mol}$ is generally conserved for each trajectory that hops.
Interestingly, for spin-irrelevant, time-reversible systems, this hopping problem can be nominally avoided using existing tricks in the literature. For instance, one can eliminate the translational and rotational component by adding electronic translational factors (ETFs) \cite{bates:1958:etf,schneiderman:1969:pr:etf,delos:1978:pra:etf,winter:1982:pra:etf,errea:1994:etf,ohrn:1994:rmp:etf,riera:1998:PRL,fatehi2012Derivative} and electronic rotational factors (ERFs) \cite{athavale2023Surface,shu2020Conservation}. After these corrections, the derivative couplings satisfy (to the first order of $m_e/M$)
\begin{align}
    \sum_I{d_{jk,\text{ETF}}^{I\alpha}} &= 0 \\
    \sum_{I,\beta,\gamma}{\epsilon_{\alpha\beta\gamma}R_{I\beta}d_{jk,\text{ETF+ERF}}^{I\gamma}} &= 0
\end{align}
Moreover, the expectation values of electronic momentum ($\ev{\hat{\mathbf{p}}}{k}$) and angular momentum ($\ev{\hat{\mathbf{l}}+\hat{\mathbf{s}}}{k}$) are zero on each adiabat (as a consequence of time-reversibility). Thus, the total linear momentum ($\mathbf{P}_{mol}$) and angular momentum  ($\mathbf{L}_{mol}$) will not change if the rescaling is done along the ETF/ERF-boosted derivative coupling directions. In other words, FSSH will conserve the total momentum and total angular momentum within a single trajectory. 

Unfortunately, however, the strategy above is not general and is not appropriate for systems with an odd number of electrons. In such a case, the expectation value of electronic angular momentum will be surface dependent (i.e., in general $\ev{\hat{\mathbf{l}}+\hat{\mathbf{s}}}{k}\ne \ev{\hat{\mathbf{l}}+\hat{\mathbf{s}}}{j}$ for $k\ne j$), so that $\mathbf{L}_{mol}$ can change during a hop even if $\mathbf{L}_{nu}$ remains constant and there is no easy way to maintain angular momentum conservation.
In the end, our feeling is that even though many of the nuances of momentum conservation can  seemingly be swept under the rug for systems with an odd number of electrons, the best semiclassical approach is to treat systems with both odd and even numbers of electrons equivalently. For instance, it is straightforward to show\cite{mead1979noncrossing} that for a system with an even number of electrons, the electronic Hamiltonian can be made strictly real valued with zero on-diagonal Berry curvature (unlike the case with an odd number of electrons). Nevertheless, as showed in Ref.~\cite{bian2021Modeling,bian2022Modeling}, the best semiclassical approach is clearly to directly treat the non-diagonal Berry forces (for a system with an even number of electrons) just as one would treat the diagonal Berry forces (for a system with an odd number of electrons). Thus, in general, one would like to do better than FSSH when it comes to linear and angular momentum conservation, which brings us to the notion of phase-space surface hopping. 

\section{Phase-Space Surface Hopping} \label{sec:pssh}

\subsection{The PSSH Algorithm} \label{sec:pssh_algorithm}

The PSSH algorithm \cite{shenvi2009Phasespace}, originally proposed by Shenvi, is one approach forward towards momentum conservation. According to PSSH, one runs normal surface hopping dynamics but with a small twist: one builds an electronic Hamiltonian that depends on both nuclear position and momentum by incorporating the derivative coupling terms explicitly into the nuclear equation of motion. The full nonadiabatic electronic Hamiltonian is given by
\begin{align}
    H_{jk}(\mathbf{R},\mathbf{P}) = V_{jk} - i\hbar\sum_{I,\alpha}{\frac{P_{I\alpha}}{M_I}{d^{I\alpha}_{jk}}} - \hbar^2\sum_{I,\alpha,l}{\frac{d^{I\alpha}_{jl}d^{I\alpha}_{lk}}{2M_I}} \label{eq:Hna}
\end{align}
By diagonalizing the Hamiltonian, generating the derivative couplings, and then re-diagonalizing the Hamiltonian in Eq.~\eqref{eq:Hna}, Shenvi argues (and has some data proving) that this dressing of the electronic states by momentum can yield some very powerful results \cite{shenvi2009Phasespace}.
In a recent paper (Ref.~\cite{wu2022phasespace}), we have argued that a similar formalism can also be valid in a totally different basis (other than adiabatic basis). More generally, in order to deliver the most insight on the nature of conservation law in PSSH-like methods, we will now consider a PSSH with an arbitrary vector-valued electronic operator $\bm{\Gamma}_{jk}$ that couples to momentum (and replaces the derivative coupling in Eq.~\eqref{eq:Hna}):
\begin{align}
    H_{jk}(\mathbf{R},\mathbf{P}) = V_{jk} - i\hbar\sum_{I,\alpha}{\frac{P_{I\alpha}}{M_I}{\Gamma^{I\alpha}_{jk}}} \label{eq:HnaGamma}
\end{align}
To make our analysis more concise, in Eq. \ref{eq:HnaGamma}, we have dropped the second derivative coupling term (the last term in Eq.~\eqref{eq:Hna}). We will show in Appendix \ref{sec:second_order}, that the inclusion of the second derivative coupling terms does not change any of the results below.

For the sake of concreteness, let us now review the PSSH algorithm that revolves around Eq.~\eqref{eq:HnaGamma}. The phase-space (PS) adiabats $\ket{n}$ are linear combinations of the selected BO states, where the coefficients are obtained by diagonalizing the nonadiabatic Hamiltonian Eq.~\eqref{eq:HnaGamma}:
\begin{align}
    \sum_k{H_{jk}(\mathbf{R},\mathbf{P})\braket{k}{n}} = E^{PS}_{n}(\mathbf{R},\mathbf{P})\braket{j}{n} \label{eq:psadiabat}
\end{align}

Like in FSSH, each trajectory in PSSH is assigned an active surface $\ket{\tilde{n}}$ and a set of amplitudes $c_n$ on different PS surfaces. These quantities are propagated by Hamilton's equation and the time-dependent Schrodinger equation:
\begin{align}
    \dot{R}_{I\alpha} &= \pdv{E_{\tilde{n}}^{PS}}{P_{I\alpha}} + \frac{P_{I\alpha}}{M_I} =  \frac{P_{I\alpha} - i\hbar\sum_{j,k}{\Gamma_{jk}^{I\alpha}\braket{\tilde{n}}{j}\braket{k}{\tilde{n}}}}{M_I} \label{eq:psshR} \\
    \dot{P}_{I\alpha} &= - \pdv{E_{\tilde{n}}^{PS}}{R_{I\alpha}} = -\sum_{j,k}{\braket{\tilde{n}}{j}\braket{k}{\tilde{n}}\nabla_{I\alpha} H_{jk}} \label{eq:psshP} \\ 
    \dot{c}_m &= -\frac{i}{\hbar}E_mc_m - \sum_{I,\alpha,n}{\left(\dot{R}_{I\alpha} \xi^{I\alpha}_{mn} - \frac{P_{I\alpha}}{M_I} \sum_{j,k}{\Gamma_{jk}^{I\alpha}\braket{m}{j}\braket{k}{n}}+ \dot{P}_{I\alpha}\tau_{mn}^{I\alpha}\right)c_n }
    \label{eq:psshc}
\end{align}
Here, 
\begin{align}
  \xi^{I\alpha}_{mn}&\equiv\bra{m}{\pdv{}{R_{I\alpha}}}\ket{n} \\  
  \tau^{I\alpha}_{mn}&\equiv\bra{m}{\pdv{}{P_{I\alpha}}}\ket{n}
\end{align}
are the total position and momentum derivative couplings (respectively) between phase-space adiabats.
A derivation of Eq.~\eqref{eq:psshc} is found in Appendix \ref{sec:eom_psshc}. Note that, according to Eq.~\eqref{eq:psshR}, PSSH dynamics always include vector potentials and Berry forces in the sense that $\mathbf{\dot{R}} \ne \mathbf{P}/\mathbf{M}$.

By defining the adiabatic density matrix of the active surface $\sigma^{[\tilde{n}]}_{jk}\equiv\sum_{j,k}{\braket{j}{\tilde{n}}\braket{\tilde{n}}{k}}$ and the transition density matrix $\sigma^{[m\to n]}_{jk} \equiv \braket{j}{n}\braket{m}{k}$, Eqs.~\eqref{eq:psshR}-\eqref{eq:psshc} can be recast as
\begin{align}
    \dot{R}_{I\alpha} &= \frac{P_{I\alpha} - i\hbar\tr[\sigma^{[\tilde{n}]} \Gamma_{I\alpha}]}{M_I} \label{eq:psshdmR} \\
    \dot{P}_{I\alpha} &= -\tr[\sigma^{[\tilde{n}]}\nabla_{I\alpha} H] = -\tr[\sigma^{[\tilde{n}]}\left(\nabla_{I\alpha} V - i\hbar\sum_{J,\delta}{\frac{P_{J\delta}}{M_J}\nabla_{I\alpha}\Gamma_{J\delta}}\right)]  \label{eq:psshdmP} \\
    \dot{c}_m &= -\frac{i}{\hbar}E_mc_m - \sum_{I,\alpha,n}{\left(\dot{R}_{I\alpha} \xi^{I\alpha}_{mn} - \frac{P_{I\alpha}}{M_I} \tr[\sigma^{[m\to n]}\Gamma_{I\alpha}]+ \dot{P}_{I\alpha}\tau_{mn}^{I\alpha}\right)c_n }
\end{align}
For a trajectory on active surface $\tilde{m}$, at each step, the hopping probability to surface $n$ is: 
\begin{align}
  g_{\tilde{m}\to n} = \max\left(2\Re\left[\sum_{I,\alpha}{(\dot{R}_{I\alpha}\xi^{I\alpha}_{\tilde{m}n}-\frac{P_{I\alpha}}{M_I}\tr[\sigma^{[\tilde{m}\to n]}\Gamma_{I\alpha}] + \dot{P}_{I\alpha}\tau^{I\alpha}_{\tilde{m}n})}\frac{c_n}{c_{\tilde{m}}}\right],0\right)  
\end{align}
Alternatively, by plugging in Eq.~\eqref{eq:psshdmR}, the hopping probability can be written as
\begin{align}
    g_{\tilde{m}\to n} &= \max\left(2\Re\left[\sum_{I,\alpha}{\left(\frac{P_{I\alpha}}{M_I}\left(\xi^{I\alpha}_{\tilde{m}n}-\tr[\sigma^{[\tilde{m}\to n]}\Gamma_{I\alpha}]\right) \right.}\right.\right. \nonumber\\
    &\qquad\left.\left.\left.+ \dot{P}_{I\alpha}\tau^{I\alpha}_{\tilde{m}n} - i\hbar\xi_{\tilde{m}n}^{I\alpha}\tr[\sigma^{[\tilde{m}]}\Gamma_{I\alpha}]\right)\frac{c_n}{c_{\tilde{m}}}\right],0\right)  \label{eq:pssh_hop}
\end{align}

Upon a successful hop, the canonical momentum ($\mathbf{P}$) is rescaled to maintain energy conservation. Unlike Shenvi's PSSH\cite{shenvi2009Phasespace} or our pseudo-diabatic PSSH\cite{wu2022phasespace}, the rescaling direction for a general Hamiltonian of the form in \eqref{eq:HnaGamma} is not  clear yet. Nevertheless, by analogy to FSSH, according to Eq.~\eqref{eq:pssh_hop}, a reasonable choice for the rescaling direction is
\begin{align}
  \bm{\lambda}_{\tilde{m}\to n} = \bm{\xi}_{\tilde{m}n}-\tr[\sigma^{[\tilde{m}\to n]}\mathbf{\Gamma}] \label{eq:pssh_rescale} 
\end{align}
Below we will analyze this solution and discuss the relevant conservation laws in Sec.~\ref{sec:pssh_rescale}.


\subsection{Conservation Laws in PSSH During Dynamics Along a Phase-Space Adiabatic Surface}
\label{sec:cons:big}
The fundamental results of this paper are as follows: PSSH dynamics will conserve the total linear momentum when moving along a given phase space adiabat if for all $\alpha,\delta,J,j,k$,
\begin{align}
    -i\hbar\sum_{I}{\Gamma^{I \alpha}_{jk}} + \mel{j}{\hat{p}_\alpha}{k} &= 0 \label{eq:GammaTcov} \\
    \sum_{I}{\nabla_{I\alpha}{\Gamma^{J\delta}_{jk}}} &= 0 \label{eq:GammagradTcov}
\end{align}
Similarly, PSSH dynamics will conserve the total angular momentum when moving along a given phase space adiabat if for all $\alpha,\delta,J,j,k$,
\begin{align}
    -i\hbar\sum_{I,\beta,\gamma}{\epsilon_{\alpha\beta\gamma}R_{I\beta}\Gamma^{I\gamma}_{jk}} + \mel{j}{\hat{l}_\alpha+\hat{s}_\alpha}{k} &= 0 \label{eq:GammaRcov}\\    
    \sum_{I,\beta,\gamma}{\epsilon_{\alpha\beta\gamma}R_{I\beta}\nabla_{I\gamma}\Gamma^{J\delta}_{jk}} + \sum_{\zeta}{\epsilon_{\alpha\delta\zeta}\Gamma^{J\zeta}_{jk}} &= 0 \label{eq:GammagradRcov}
\end{align}
Furthermore, we show in Appendix \ref{sec:pssh_symdrvgrad} that, provided we use a basis satisfying Eqs.~\eqref{eq:Tcov} and \eqref{eq:Rcov}, Shenvi's PSSH (i.e. Eq.~\eqref{eq:HnaGamma} where $\bm{\Gamma}=\mathbf{d}$ is used) will conserve both linear and angular momentum.

\subsubsection{The Linear Momentum} \label{sec:pssh_ptot}

For a trajectory propagated on phase-space adiabat $\tilde{n}$, the change in the total molecular linear momentum $\mathbf{P}_{mol}$  is (using. Eq.~\eqref{eq:psshdmR}):
\begin{align}
    \dv{P_{mol,\alpha}}{t} &= \dv{t} \left(\sum_I{M_I\dot{R}_{I\alpha}} + \tr[\sigma^{[\tilde{n}]} \hat{p}_\alpha]\right) \nonumber\\
    &= \sum_I{\dot{P}_{I\alpha}} + \dv{t}\tr[\sigma^{[\tilde{n}]}\left(-i\hbar\sum_I{\Gamma_{I\alpha}} + \hat{p}_\alpha\right)]
\end{align}
According to Eq.~\eqref{eq:GammaTcov}, the last term is zero, which implies $\dot{P}_{mol,\alpha} = \sum_I{\dot{P}_{I\alpha}} = -\sum_I{\tr[\sigma^{[\tilde{n}]} \nabla_{I\alpha}H]}$. Plugging in Eq.~\eqref{eq:psshdmP}, we find
\begin{align}
    \dv{P_{mol,\alpha}}{t} = &-\sum_I{\tr[\sigma^{[\tilde{n}]} \nabla_{I\alpha}H]} \nonumber\\
    &=-\sum_I{\tr[\sigma^{[\tilde{n}]}\left(\nabla_{I\alpha} V - i\hbar\sum_{J,\delta}{\frac{P_{J\delta}}{M_J}\nabla_{I\alpha} \Gamma_{J\delta}}\right)]} \label{eq:dPmol_pssh2}
\end{align}
According to Eqs.~\eqref{eq:bogradTinv} and \eqref{eq:GammagradTcov}, $\sum_I{\nabla_{I\alpha}V_{jk}}=0$ and $\sum_I{\nabla_{I\alpha}\Gamma^{J\delta}_{jk}}=0$ for all $j,k,J,\delta$. Therefore Eq.~\eqref{eq:dPmol_pssh2} evaluates to zero. As a result, PSSH conserves the total linear momentum when a trajectory is propagated along a phase-space adiabat.

\subsubsection{The Angular Momentum} \label{sec:pssh_ltot}

According to the definition of the total molecular angular momentum $\mathbf{L}_{mol}$ in Eq.~\eqref{eq:Lmol}, for a trajectory propagating on phase-space adiabat $\tilde{n}$, the change in the total angular momentum is (again using Eq.~\eqref{eq:psshdmR}):
\begin{align}
    \dv{L_{mol,\alpha}}{t} &= \dv{t} \left(\sum_{I,\beta,\gamma}{\epsilon_{\alpha\beta\gamma}M_IR_{I\beta}\dot{R}_{I\gamma}} + \tr[\sigma^{[\tilde{n}]}(\hat{l}_\alpha + \hat{s}_\alpha)]\right) \nonumber\\
    &= \dv{t} \sum_{I,\beta,\gamma}{\epsilon_{\alpha\beta\gamma}R_{I\beta}P_{I\gamma}} + \dv{t}\tr[\sigma^{[\tilde{n}]}\left(-i\hbar\sum_{I,\beta,\gamma}{\epsilon_{\alpha\beta\gamma}R_{I\beta}\Gamma_{I\gamma}} + \hat{l}_\alpha + \hat{s}_\alpha\right)]
\end{align}
According to Eq.~\eqref{eq:GammaRcov}, the second term is zero, and therefore
\begin{align}
    \dv{L_{mol,\alpha}}{t} = \dv{t} \sum_{I,\beta,\gamma}{\epsilon_{\alpha\beta\gamma}R_{I\beta}P_{I\gamma}} &= \left(\sum_{I,\beta,\gamma}{\epsilon_{\alpha\beta\gamma}(\dot{R}_{I\beta}P_{I\gamma} + R_{I\beta}\dot{P}_{I\gamma})}\right)
\end{align}
Plugging in Eq.~\eqref{eq:psshdmR} for $\dot{\mathbf{R}}$ and Eq.~\eqref{eq:psshdmP} for $\dot{\mathbf{P}}$, we find
\begin{align}
    \dv{L_{mol,\alpha}}{t}
    &=  \sum_{I,\beta,\gamma} {\epsilon_{\alpha\beta\gamma}\left(\frac{P_{I\beta}P_{I\gamma}}{M_I} -i\hbar\tr[\sigma^{[\tilde{n}]} \Gamma_{I\beta}]\frac{P_{I\gamma}}{M_I}\right.} \nonumber\\
    &\qquad\left.-R_{I\beta}\tr[\sigma^{[\tilde{n}]}\left(\nabla_{I\gamma} V - i\hbar\sum_{J,\delta}{\frac{P_{J\delta}}{M_J}\nabla_{I\gamma}\Gamma_{J\delta}}\right)]
 \right) \label{eq:dJmol_pssh2}
\end{align}
In Eq.~\eqref{eq:dJmol_pssh2}, the first term is zero since it is a cross product between a vector and itself, and according to Eq.~\eqref{eq:bogradRinv}, the $\nabla_{I\gamma}V$ term is also zero. The remaining terms are
\begin{align}
    \dv{L_{mol,\alpha}}{t} 
    &=-i\hbar\sum_{I,\beta,\gamma}{\epsilon_{\alpha\beta\gamma}\tr[\sigma^{[\tilde{n}]}\left(\Gamma_{I\beta}\frac{P_{I\gamma}}{M_I} - R_{I\beta}\sum_{J,\delta}{\frac{P_{J\delta}}{M_J}\nabla_{I\gamma}\Gamma_{J\delta}}\right)]} \label{eq:dJmol_pssh2.5}
\end{align}
Plugging in Eq.~\eqref{eq:GammagradRcov}, we find
\begin{align}
    \dv{L_{mol,\alpha}}{t} 
    &=-i\hbar\sum_{I,\beta,\gamma}{\epsilon_{\alpha\beta\gamma}\tr[\sigma^{[\tilde{n}]}\Gamma_{I\beta}\frac{P_{I\gamma}}{M_I}]} - i\hbar\sum_{J,\delta,\zeta}{\epsilon_{\alpha\delta\zeta}\tr[\sigma^{[\tilde{n}]}\frac{P_{J\delta}}{M_J}\Gamma_{J\zeta}]}
    \label{eq:dJmol_pssh3}
\end{align}
Replacing the dummy indices $\delta\to\gamma,\zeta\to\beta,J\to I$, and utilizing the antisymmetric property of the Levi-Civita symbol, Eq.~\eqref{eq:dJmol_pssh3} becomes
\begin{align}
    \dv{L_{mol,\alpha}}{t} 
    &=-i\hbar\sum_{I,\beta,\gamma}{\epsilon_{\alpha\beta\gamma}\tr[\sigma^{[\tilde{n}]}\Gamma_{I\beta}\frac{P_{I\gamma}}{M_I}]} - i\hbar\sum_{I,\beta,\gamma}{\epsilon_{\alpha\gamma\beta}\tr[\sigma^{[\tilde{n}]}\frac{P_{I\gamma}}{M_I}\Gamma_{I\beta}]} = 0 \label{eq:dJmol_pssh4}
\end{align}
Thus, PSSH conserves the total molecular angular momentum when a trajectory is propagated along a phase-space adiabat.

\subsection{Conservation Laws in PSSH In the Course of a Hop} \label{sec:pssh_rescale}

Here we discuss the effect of momentum rescaling (according to Eq.~\eqref{eq:pssh_rescale})
as far as the relevant conservation laws.
 We assume that the hop is from phase-space adiabat $m$ to $n$.

Before we begin our discussion, one point must be emphasized: Although $\bm{\lambda}_{m\to n}$ defines a direction in coordinate space, this vector cannot be used in a naive black-box fashion, as both $\bm{\xi}$ and $\bm{\Gamma}$ in Eq.~\eqref{eq:pssh_rescale} are usually complex-valued and gauge-dependent. In practice, a phase factor is required if we wish to map the direction $\bm{\lambda}_{m\to n}$ onto a real-vector in the Cartesian space. Therefore, the general expression for the rescaled canonical momentum should read
\begin{align}
    \Delta\mathbf{P}_{m\to n} = \hbar\eta\Re[\bm{\lambda}_{m\to n} e^{i\phi}]
\end{align}
where $\eta$ is a unitless (one dimensional) rescaling amplitude and $e^{i\phi}$ is the relevant phase factor. In Ref.~\cite{wu2023quantumclassical}, we have shown that from the quantum-classical Liouville equation (QCLE), a good way to choose such phase factor is ${\mathbf{P}\cdot\bm{\lambda}_{m\to n}^*}/{\abs{\mathbf{P}\cdot\bm{\lambda}_{m\to n}^*}}$. Below, we will discuss this choice of phase and others.


In order to proceed any further, it will be necessary to evaluate the matrix elements of $\bm{\lambda}_{m\to n} = \bm{\xi}_{mn}-\tr[\sigma^{[m\to n]}\mathbf{\Gamma}]$.
Now, $\bm{\xi}_{mn}$ can be better understood by repeatedly inserting  resolutions of the identity, $\sum_k{\ket{k}\bra{k}}$ and $\sum_j{\ket{j}\bra{j}}$:
\begin{align}
    \xi^{I\alpha}_{mn} &\equiv \bra{m}\nabla_{I\alpha}\ket{n} = \sum_k{ \bra{m}\nabla_{I\alpha}(\ket{k}\bra{k}\ket{n}) }\nonumber\\
    &= \sum_k{\braket{m}{\nabla_{I\alpha} k}\braket{k}{n}} + \sum_k{\braket{m}{k}\nabla_{I\alpha}(\braket{k}{n})} \nonumber\\
    &= \sum_{j,k}{\braket{m}{j}\braket{j}{\nabla_{I\alpha} k}\braket{k}{n}} + \sum_k{\braket{m}{k}\nabla_{I\alpha}(\braket{k}{n})} \label{eq:xi}
\end{align}
In the first term in Eq.~\eqref{eq:xi}, $\braket{j}{\nabla_{I\alpha}k} = d^{I\alpha}_{jk}$ is just the normal adiabatic derivative coupling. The second term in Eq.~\eqref{eq:xi} arises from the rotation from adiabats to the phase-space adiabats, and can be evaluated by the Hellmann-Feynman theorem (note here we assume phase-space adiabats are non-degenerate, which is reasonable since the matrix $\bm{\Gamma}_{jk}$ is generally dense):
\begin{align}
    &\sum_k{\braket{m}{k}\nabla_{I\alpha}(\braket{k}{n})} \nonumber\\
    &\qquad = \frac{1}{E_n^{PS}-E_m^{PS}}\sum_{j,k}{\braket{m}{j}(\nabla_{I\alpha}V_{jk} - i\hbar \sum_{J,\delta}{\frac{P_{J\delta}}{M_J}\nabla_{I\alpha}\Gamma_{jk}^{J\delta}})\braket{k}{n}} \label{eq:superdc_hf}
\end{align}
Altogether, if we plug Eqs.~\eqref{eq:superdc_hf} and \eqref{eq:xi} into Eq.~\eqref{eq:pssh_rescale}, and substitute the definition of $\sigma^{[m\to n]}$, we arrive at
\begin{align}
    \lambda^{I\alpha}_{m\to n} = \tr[\sigma^{[m\to n]}\left(d_{I\alpha} - \Gamma_{I\alpha} + \frac{\nabla_{I\alpha} V - i\hbar\sum_{J,\delta}{ \frac{P_{J\delta}}{M_J}\nabla_{I\alpha}\Gamma_{J\delta}}}{E_n^{PS}-E_m^{PS}}\right)] \label{eq:lambda_exp}
\end{align}

Finally, when considering both linear and angular momentum below, it will be helpful to 
evaluate the change in the nuclear kinetic momentum component after a hop. According to Eq.~\eqref{eq:psshdmR}, we find this quantity is: 
\begin{align}
    M_I\Delta\dot{R}_{I\alpha}^{m\to n} = \hbar\eta\Re[\lambda^{I\alpha}_{m\to n}e^{i\phi}] - i\hbar\tr[\Gamma_{I\alpha}(\sigma^{[n]}-\sigma^{[m]})]
\end{align}

\subsubsection{Energy Conservation}

Unlike FSSH, the surfaces in PSSH depend on nuclear momentum and will change upon momentum rescaling. Therefore, in order to satisfy energy conservation exactly, one generally cannot solve for the rescaling amplitude $\eta$ analytically, but rather one must solve a\ self-consistent equation:
\begin{align}
    E_n^{PS}(\mathbf{R},\mathbf{P}+\hbar\eta\Re[\bm{\lambda}_{m\to n}e^{i\phi}]) = E_m^{PS}(\mathbf{R},\mathbf{P})
\end{align}
where the functional $E_m^{PS}(\mathbf{R},\mathbf{P})$ is given by Eq.~\eqref{eq:psadiabat}.

\subsubsection{The Linear Momentum in Rescaling}
The change of molecular linear momentum is given by
\begin{align}
    \Delta P_{mol,\alpha}^{m\to n} &= \Delta P_{nu,\alpha}^{m\to n} + \Delta P_{el,\alpha}^{m\to n} \nonumber\\
    &= \sum_I{M_I\Delta \dot{R}^{m\to n}_{I\alpha}} + \ev{\hat{p}_\alpha}{n} - \ev{\hat{p}_\alpha}{m} \nonumber\\
    &= \hbar\eta\sum_I{\Re[\lambda^{m\to n}_{I\alpha} e^{i\phi}]} + \tr[(\sigma^{[n]}-\sigma^{[m]})\left(\hat{p}_\alpha - i\hbar\sum_I{\Gamma_{I\alpha}}\right)] \label{eq:deltaPmol} 
\end{align}
According to Eqs.~\eqref{eq:GammaTcov}, the second term of Eq.~\eqref{eq:deltaPmol} is zero. If we substitute Eq.~\eqref{eq:lambda_exp} for the first term, we find
\begin{align}
    \Delta P^{m\to n}_{mol,\alpha}
    &= \hbar\eta\Re\left[e^{i\phi}\tr[\sigma^{[m\to n]}\sum_I{\left(d_{I\alpha} - \Gamma_{I\alpha} + \frac{\nabla_{I\alpha} V - i\hbar\sum_{J,\delta}{ \frac{P_{J\delta}}{M_J}\nabla_{I\alpha}\Gamma_{J\delta}}}{E_n^{PS}-E_m^{PS}}\right)}]\right] \label{eq:deltaPmol_pssh}
\end{align}
Let us now examine the individual terms in Eq.~\eqref{eq:deltaPmol_pssh}. According to Eqs.~\eqref{eq:dcTcov}, \eqref{eq:bogradTinv}, we have
\begin{align}
    \sum_I{(d_{jk}^{I\alpha} - \Gamma_{jk}^{I\alpha})} = -\frac{i}{\hbar}(\mel{j}{\hat{p}_\alpha}{k} -\mel{j}{\hat{p}_\alpha}{k}) &= 0
\end{align}
According to Eqs.~\eqref{eq:GammaTcov} and \eqref{eq:GammagradTcov}, we also have
\begin{align}
    \sum_I{(\nabla_{I\alpha}V_{jk} - i\hbar \sum_{J,\delta}{ \frac{P_{J\delta}}{M_J}\nabla_{I\alpha}\Gamma_{J\delta}})} = (0-i\hbar\sum_{J,\delta}{\frac{P_{J\delta}}{M_J}\cdot 0} ) &= 0
\end{align}
Therefore, every term in Eq.~\eqref{eq:deltaPmol_pssh} evaluates to zero, which indicates that the total molecular linear momentum does not change during  momentum rescaling, regardless of the phase factor used.

\subsubsection{The Angular Momentum in Rescaling}
The change of molecular angular momentum is given by
\begin{align}
    \Delta L^{m\to n}_{mol,\alpha} &= \Delta L^{m\to n}_{nu,\alpha} + \Delta L^{m\to n}_{el,\alpha} + \Delta S^{m\to n}_{el,\alpha} \nonumber\\
    &= \sum_{I,\beta,\gamma}{\epsilon_{\alpha\beta\gamma}M_IR_{I\beta}\Delta\dot{R}^{m\to n}_{I\gamma}} + \ev{\hat{l}_\alpha+\hat{s}_\alpha}{n} - \ev{\hat{l}_\alpha+\hat{s}_\alpha}{m} \nonumber\\
    &= \hbar\eta\sum_{I,\beta,\gamma}{\epsilon_{\alpha\beta\gamma}\Re[R_{I\beta}\lambda^{m\to n}_{I\gamma}e^{i\phi}]} \nonumber\\
    &\qquad- \tr[(\sigma^{[n]}-\sigma^{[m]})\left(\hat{l}_\alpha + \hat{s}_\alpha - i\hbar\sum_{I,\beta,\gamma}{\epsilon_{\alpha\beta\gamma}R_{I\beta}\Gamma_{I\gamma}}\right)] \label{eq:deltaLmol}
\end{align}
Similar to the case of linear momentum, according to Eq.~\eqref{eq:GammaRcov}, the second term of Eq.~\eqref{eq:deltaLmol} is zero. If we substitute Eq.~\eqref{eq:lambda_exp} for the first term, we find
\begin{align}
    \Delta L^{m\to n}_{mol,\alpha} &= \hbar\eta\Re\left[e^{i\phi}\tr\left[\sigma^{[m\to n]}\sum_{I,\beta,\gamma}{\epsilon_{\alpha\beta\gamma}R_{I\beta}\left(d_{I\gamma} - \Gamma_{I\gamma} \right.}\right.\right. \nonumber\\
    &\qquad+ \left.\left.\left. \frac{\nabla_{I\gamma} V - i\hbar\sum_{J,\delta}{\frac{P_{J\delta}}{M_J}\nabla_{I\gamma}\Gamma_{J\delta}}}{E_n^{PS}-E_m^{PS}}\right)\right]\right] \label{eq:deltaLmol_pssh}
\end{align}
Again, we examine the individual terms in Eq.~\eqref{eq:deltaLmol_pssh}. According to Eqs.~\eqref{eq:dcRcov} and \eqref{eq:bogradRinv}, we have
\begin{align}
    &\sum_{I,\beta,\gamma}{\epsilon_{\alpha\beta\gamma}R_{I\beta}(d_{jk}^{I\gamma} - \Gamma_{jk}^{I\gamma})} = -\frac{i}{\hbar}(\mel{j}{\hat{l}_\alpha+\hat{s}_\alpha}{k} - \mel{j}{\hat{l}_\alpha+\hat{s}_\alpha}{k}) = 0
\end{align}
According to \eqref{eq:GammaRcov} and \eqref{eq:GammagradRcov}, we also  have
\begin{align}
    &\sum_{I,\beta,\gamma}{\epsilon_{\alpha\beta\gamma}R_{I\beta}\left(\nabla_{I\gamma}V_{jk} - i\hbar \sum_{J,\delta}{\nabla_{I\gamma}\Gamma_{jk}^{J\delta}}\frac{P_{J\delta}}{M_J}\right)} \nonumber\\
    &\qquad\qquad = 0+ i\hbar\sum_{J,\delta,\zeta}{\epsilon_{\alpha\delta\zeta}\Gamma_{jk}^{J\zeta}\frac{P_{J\delta}}{M_J}} = i\hbar\sum_{I,\beta,\gamma}{\epsilon_{\alpha\beta\gamma}\frac{P_{I\beta}}{M_I}\Gamma_{I\gamma}}
\end{align}
Therefore, the total angular momentum change during the momentum rescaling is
\begin{align}
    \Delta L_{mol,\alpha}
    &=\hbar\eta\Re\left[e^{i\phi}\frac{i\hbar}{E_n^{PS}-E_m^{PS}}\tr[\sigma^{[m\to n]} \sum_{I,\beta,\gamma}{\epsilon_{\alpha\beta\gamma}\frac{P_{I\beta}\Gamma_{I\gamma}}{M_I}}]\right] \nonumber\\
    &= -\frac{\hbar^2\eta}{E_n^{PS}-E_m^{PS}}\Im\left[e^{i\phi}\tr[\sigma^{[m\to n]} \sum_{I,\beta,\gamma}{\epsilon_{\alpha\beta\gamma}\frac{P_{I\beta}\Gamma_{I\gamma}}{M_I}}]\right] \label{eq:pssh_rescale_Lmol_final}
\end{align}
Eq.~\eqref{eq:pssh_rescale_Lmol_final} cannot be further simplified. Thus, the momentum rescaling scheme proposed in Eq.~\eqref{eq:pssh_rescale} will generally (and unfortunately) bring about a nonzero change in the total molecular angular momentum. That being said, it is crucial to emphasize that the magnitude of the change in angular momentum scales as $\hbar^2$ in Eq.~\eqref{eq:pssh_rescale_Lmol_final} -- whereas the corresponding change would scale as $\hbar$ in FSSH. Empirically, we have found in a few test cases (unpublished) that this error is usually very small.

\subsubsection{Restoring the Exact Conservation Laws}
Finally, we note that, if one is determined to satisfy conservation of linear momentum and angular momentum exactly, there is one very straightforward path. Namely, one can pick the rescaling direction to be: 
\begin{align}
\label{exact_rescale}
\bm{\lambda}_{m\to n} \equiv \tr[\sigma^{[m\to n]}(\mathbf{d} - \bm{\Gamma})]
\end{align}
In other words, one merely drops the second term in Eq.~\eqref{eq:xi}. Admittedly, such a rescaling does not reduce to Shenvi's PSSH algorithm where $\bm{\Gamma}=\mathbf{d}$ (because the rescaling direction would be undefined in such a case). Nevertheless, if one can choose a relatively smooth and small $\bm{\Gamma}$, the algorithm will be well-defined and the rescaling direction should be close the original PSSH approach as well.

Lastly, as far as the phase $e^{i\phi}$ is concerned, one might be tempted to choose $\phi$ such that $\abs{\Delta\mathbf{L}_{mol}}$ is minimized (and the change in angular momentum is even further reduced). After some experience with PSSH, however, our feeling is that this is not a productive path forward. In general, following Ref.~\cite{miao2019extension}, it seems best to simply choose $\phi$ such that $\abs{\Re[\bm{\lambda}e^{i\phi}]}$ is maximal.

\section{Conclusion} \label{sec:conclusion}

In this paper, we have analyzed the conservation of linear and angular momentum within two different surface hopping methods: The standard, fewest switches surface hopping (FSSH) by Tully, and the phase-space surface hopping (PSSH) approach of Shenvi. In a separate paper, we have recently analyzed the relevant Ehrenfest dynamics, and the reader should also see Ref.~\citenum{li2022Energy} for a relevant discussion in terms of exact factorization.

For FSSH, the electronic Hamiltonian depends only on nuclear position. In such a case, neither the total linear nor the total angular momentum of a trajectory is conserved if we account for electronic momentum or angular momentum.
The FSSH algorithm conserves only the {\em nuclear} momenta. Moreover, if we hop in the direction of the derivative coupling, the resulting momentum rescaling will break both linear and angular momentum conservation. Patches like ETFs and ERFs (e.g. those developed by Shu {\em et al}\cite{shu2020Conservation} and our group \cite{athavale2023Surface}) can restore the conservation of the nuclear momenta, but the algorithm still will not recover the correct total momenta. The latter scenario should be most problematic in the presence of degeneracy (e.g. a Kramers' spin doublet) when the surfaces are time-irreversible.

For PSSH, the electronic Hamiltonian depends on both nuclear position ($\mathbf{R}$) and nuclear momentum ($\mathbf{P}$). We imagine adding a term $\bm{\Gamma}\cdot\mathbf{P}$ where $\bm{\Gamma}$ is a matrix that depends on position ($\mathbf{R}$) and is to be determined. We find that such a generalized PSSH conserves both the linear and the angular momentum of a trajectory during propagation provided that $\bm{\Gamma}$ satisfies the symmetry constraints in Eqs.~\eqref{eq:GammaTcov}-\eqref{eq:GammagradRcov}; all Berry forces are automatically included in PSSH.
Moreover, if the momentum rescaling direction is chosen as in Eq. (\ref{exact_rescale}), the resulting PSSH algorithm exactly conserves linear and angular momentum during a hop; if the momentum rescaling is chosen as in  Eq.~\eqref{eq:pssh_rescale}, the algorithm nearly conserves angular momentum change during a hop, but not exactly -- the resulting error should be proportional to $\hbar^2$. Eqs.~\eqref{eq:GammaTcov}-\eqref{eq:GammagradRcov} are satisfied if we substitute $\bm{\Gamma} = \mathbf{d}$ (the actual derivative coupling), confirming that  Shenvi's original adiabatic PSSH does maintain linear and angular momentum conservation.

The most important next step in this research is how to choose $\bm{\Gamma}$. There are several reasons that one should fear setting $\bm{\Gamma}=\mathbf{d}$ (Shenvi's algorithm). First, as shown by Gherib {\em et al}, the resulting method fails near conical intersections because of the divergence of $\mathbf{d}$ \cite{gherib2016inclusion}; one wants to use PSSH to fix up the Born-Oppenheimer approximation far from a crossing but standard surface hopping already works well near a crossing \cite{izmaylov:2015:fssh_ci_whysowell} and one does not want a correction that actually makes the results worse. Second, in practice, one will need to differentiate the electronic Hamiltonian for dynamics and differentiating the derivative coupling will be extremely expensive. Third, in the case where one works with SOC and an odd number of electrons, the derivative coupling is not well-defined so that the resulting PSSH algorithm would be gauge dependent. For all of these reasons, it is quite logical to search for and explore different possible $\bm{\Gamma}$ matrices in the future. While one can certainly ``guess'' the correct $\bm{\Gamma}$ operators for some model problems \cite{wu2022phasespace,bian2022Modeling}, the optimal choice of $\bm{\Gamma}$ in general remains an important open question. In publishing this paper, our hope is that the theory community will now actively pursue this goal.

Looking forward, provided one can isolate meaningful, physically based $\Gamma$ matrix elements, it seems very possible we will be able to explore the very rich intersection of nonadiabatic dynamics and spintronics, ideally using {\em ab initio} electronic structure theory, in the near future.

\section*{Acknowledgments}
This material is based on the work supported by the National Science Foundation under Grant No.~CHE-2102402.


\appendix

\section{Gauge Conditions for Adiabats} \label{sec:adiabat_cov}

In this section we provide a brief discussion of Eqs.~\eqref{eq:Tcov} and \eqref{eq:Rcov}. For non-degenerate adiabats, it can be shown rigorously that $\mel{j}{\mathcal{P}_\alpha + \hat{p}_\alpha}{k} = 0$ and $\mel{j}{\mathcal{L}_\alpha + \hat{l}_\alpha + \hat{s}_\alpha}{k} = 0$ for any $j\ne k$. Here is a proof for translation:
\begin{align}
    \mel{j}{\mathcal{P}_\alpha + \hat{p}_\alpha}{k} &= \frac{\mel{j}{E_j(\mathcal{P}_\alpha + \hat{p}_\alpha) - (\mathcal{P}_\alpha + \hat{p}_\alpha)E_k}{k}}{E_j-E_k} \nonumber\\
    &= \frac{\mel{j}{[V,\mathcal{P}_\alpha + \hat{p}_\alpha]}{k}}{E_j-E_k} = 0 \label{eq:Tcov_proof}
\end{align}
In the last equality of Eq.~\eqref{eq:Tcov_proof} we have used Eq.~\eqref{eq:VTinv}. A similar proof holds for rotation. 
Accordingly, since  the adiabats form a complete basis, enforcing Eqs.~\eqref{eq:Tcov} and \eqref{eq:Rcov} is really just a matter of phase conventions for the case $j=k$.  These phase conventions are usually discussed in the context of the on-diagonal derivative coupling $d_{kk}$:\begin{align}
    0 &= \mel{k}{\mathcal{P}_\alpha + \hat{p}_\alpha}{k} = -\frac{i}{\hbar}\sum_I{d_{kk}^{I\alpha}} + \mel{k}{\hat{p}_\alpha}{k} \label{eq:dcTcov_diag} \\
    0 &= \mel{k}{\mathcal{L}_\alpha + \hat{l}_\alpha + \hat{s}_\alpha}{k} = -\frac{i}{\hbar}\sum_{I,\beta,\gamma}{\epsilon_{\alpha\beta\gamma}R_{I\beta}d_{kk}^{I\gamma}} + \mel{k}{\hat{l}_\alpha + \hat{s}_\alpha}{k} \label{eq:dcRcov_diag}
\end{align}
In words, Eqs.~\eqref{eq:dcTcov_diag} and \eqref{eq:dcRcov_diag} indicates that the phase of adiabats should be chosen such that the translational and rotational constraints are met. The phase choosing procedure is detailed in Ref.~\cite{littlejohn2023Representation} and Eqs.~\eqref{eq:Tcov} and \eqref{eq:Rcov} can be satisfied for states even when there is degeneracy.

\section{Equation of Motion for the Amplitudes in PSSH} \label{sec:eom_psshc}

Here we provide a derivation of the PSSH equation of motion for the amplitudes (Eq.~\eqref{eq:psshc}).
The time-dependent Schrodinger equation reads
\begin{align}
    \pdv{\ket{\psi}}{t} = -\frac{i}{\hbar}\hat{V}\ket{\psi} \label{eq:tdse}
\end{align}
Now, for the amplitude in PSSH, $c_m=\braket{m}{\psi}$, we have
\begin{align}
    \pdv{c_m}{t} = \pdv{\bra{m}}{t}\ket{\psi} + \bra{m}\pdv{\ket{\psi}}{t}
\end{align}
Using the chain rule of derivative for the first term, and substituting Eq.~\eqref{eq:tdse} for the second term, we find
\begin{align}
    \pdv{c_m}{t} = \sum_{I,\alpha}{\dot{R}_{I\alpha}\pdv{\bra{m}}{R_{I\alpha}}\ket{\psi}} + \sum_{I,\alpha}{\dot{P}_{I\alpha}\pdv{\bra{m}}{P_{I\alpha}}\ket{\psi}} - \frac{i}{\hbar}\mel{m}{\hat{V}}{\psi}
\end{align}
Inserting the resolution of identity $\sum_n \ket{n}\bra{n}$, we further find
\begin{align}
    \pdv{c_m}{t} &= \sum_{I,\alpha,n}{\dot{R}_{I\alpha}\pdv{\bra{m}}{R_{I\alpha}}\ket{n}\braket{n}{\psi}} + \sum_{I,\alpha,n}{\dot{P}_{I\alpha}\pdv{\bra{m}}{P_{I\alpha}}\ket{n}\braket{n}{\psi}} - \frac{i}{\hbar}\sum_n{\mel{m}{\hat{V}}{n}\braket{n}{\psi}} \nonumber\\
    &= -\sum_{I,\alpha,n}{\dot{R}_{I\alpha}\mel{m}{\pdv{R_{I\alpha}}}{n}c_n} - \sum_{I,\alpha,n}{\dot{P}_{I\alpha}\mel{m}{\pdv{P_{I\alpha}}}{n}c_n} - \frac{i}{\hbar}\sum_n{V_{mn}c_n} \label{eq:psshc_drv1}
\end{align}
Since $m,n$ are phase-space adiabats, according to Eq.~\eqref{eq:psadiabat}, we can gather matrix elements:
\begin{align}
    V_{mn}=E_m\delta_{mn} + i\hbar\sum_{I,\alpha,j,k}{\Gamma_{jk}^{I\alpha}\frac{P_{I\alpha}}{M_I}\braket{m}{j}\braket{k}{n}}
\end{align}
If we substitute this expression into Eq.~\eqref{eq:psshc_drv1}, we finally have
\begin{align}
    \pdv{c_m}{t} &= -\sum_{I,\alpha,n}{\left(\dot{R}_{I\alpha}\mel{m}{\pdv{R_{I\alpha}}}{n} - \frac{P_{I\alpha}}{M_I}\Gamma_{jk}^{I\alpha}\braket{m}{j}\braket{k}{n}\right) c_n} \nonumber\\
    &\qquad- \sum_{I,\alpha,n}{\dot{P}_{I\alpha}\mel{m}{\pdv{P_{I\alpha}}}{n}c_n} - \frac{i}{\hbar}E_m c_m
\end{align}
which is just Eq.~\eqref{eq:psshc}.

\section{Transformation Properties of the Gradients of The Derivative Couplings} \label{sec:pssh_symdrvgrad}

Here we will show that for a basis set $\ket{j},\ket{k}$ that satisfies Eqs.~\eqref{eq:Tcov} and \eqref{eq:Rcov}, Eq.~\eqref{eq:GammagradTcov} and \eqref{eq:GammagradRcov} will automatically hold by replacing $\bm{\Gamma}_{jk}$ with the derivative couplings $\mathbf{d}_{jk}$.

First, let us show that Eq.~\eqref{eq:GammagradTcov} holds for $\bm{\Gamma}_{jk} = \mathbf{d}_{jk}$. Expanding the LHS, we have
\begin{align}
    \sum_I{\nabla_{I\alpha}d^{J\delta}_{jk}} &= \sum_I{\left(\braket{\nabla_{I\alpha} j}{\nabla_{J\delta} k} + \bra{j}\nabla_{I\alpha}(\ket{\nabla_{J\delta} k})\right)} \nonumber\\
    &= \sum_I{\left(\braket{\nabla_{I\alpha} j}{\nabla_{J\delta} k} + \bra{j}\nabla_{{J\delta}}(\ket{\nabla_{I\alpha} k})\right)} \nonumber\\
    &\equiv \frac{i}{\hbar}(-\braket{\mathcal{P}_\alpha j}{\nabla_{J\delta} k} + \bra{j}\nabla_{J\delta}(\ket{\mathcal{P}_\alpha k}) \label{eq:dcgradTcov_drv1}
\end{align}
Utilizing Eq.~\eqref{eq:Tcov}, we find
\begin{align}
    \sum_I{\nabla_{I\alpha}d^{J\delta}_{jk}}
    &= \frac{i}{\hbar}(\braket{\hat{p}_\alpha j}{\nabla_{J\delta} k} - \braket{j}{ \nabla_{J\delta} \hat{p}_\alpha k}) \nonumber\\
    &= \frac{i}{\hbar}\mel{j}{[\hat{p}_\alpha,\nabla_{J\delta}]}{k}
\end{align}
Since $\hat{p}_\alpha$ is an electronic operator that does not depend on the nuclear position, the expression evaluates to zero.

Second, let us show that Eq.~\eqref{eq:GammagradRcov} holds for $\bm{\Gamma}_{jk} = \mathbf{d}_{jk}$. Expanding the LHS, we find
\begin{align}
    \sum_{I,\beta,\gamma}{\epsilon_{\alpha\beta\gamma}R_{I\beta}\nabla_{I\gamma}d^{J\delta}_{jk}}& = \epsilon_{\alpha\beta\gamma}\sum_I{R_{I\beta}(\braket{\nabla_{I\gamma}j}{\nabla_{J\delta}k} + \bra{j}\nabla_{I\gamma}(\ket{\nabla_{J\delta}k}))} \label{eq:dcgradRcov_drv1}
\end{align}
For the second term in the RHS of \eqref{eq:dcgradRcov_drv1}, by the chain rule of derivative, we arrive at
\begin{align}
    \sum_I{R_{I\beta}\bra{j}\nabla_{I\gamma}(\ket{\nabla_{J\delta}k})} &= \sum_I{\bra{j}\nabla_{J\delta}(R_{I\beta}\nabla_{I\gamma}\ket{k})} - \bra{j}\left[\nabla_{J\delta},\sum_I{R_{I\beta}\nabla_{I\gamma}} \right]\ket{k} \nonumber\\
    &= \sum_I{\bra{j}\nabla_{J\delta}(R_{I\beta}\nabla_{I\gamma}\ket{k})} - \delta_{IJ}\delta_{\beta\delta}\mel{j}{\nabla_{I\gamma}}{k}
\end{align}
If we plug in Eq.~\eqref{eq:dcgradRcov_drv1}, we find
\begin{align}
    \sum_{I,\beta,\gamma}{\epsilon_{\alpha\beta\gamma}R_{I\beta}\nabla_{I\gamma}d^{J\delta}_{jk}}& = \sum_{I,\beta,\gamma}{\epsilon_{\alpha\beta\gamma}R_{I\beta}\braket{\nabla_{I\gamma}j}{\nabla_{J\delta}k}} \nonumber\\
    &\qquad+ \sum_{I,\beta,\gamma}{\epsilon_{\alpha\beta\gamma}\bra{j}\nabla_{J\delta}(R_{I\beta}\ket{\nabla_{I\gamma}k})} - \sum_{\gamma}{\epsilon_{\alpha\delta\gamma}\bra{j}\nabla_{J\gamma}\ket{k}}
    \nonumber\\
    &\equiv \frac{i}{\hbar}(-\braket{\mathcal{L}_\alpha j}{\nabla_{J\delta} k} + \braket{j}{\mathcal{L}_\alpha \nabla_{J\delta} k}) - \sum_{\gamma}{\epsilon_{\alpha\delta\gamma}d_{jk}^{J\gamma}}
\end{align}
where $\mathcal{L}$ was defined in Eq.~\eqref{eq:Lop}. If we now plug in Eq.~\eqref{eq:Rcov}, the result is
\begin{align}
    \sum_{I,\beta,\gamma}{\epsilon_{\alpha\beta\gamma}\nabla_{I\gamma}d^{J\beta}_{jk}} &= \frac{i}{\hbar}\left(\braket{(\hat{l}_\alpha + \hat{s}_\alpha) j}{\nabla_{J\delta} k}\right. \nonumber\\
    &\qquad\left.- \braket{j}{\nabla_{J\delta}(\hat{l}_\alpha + \hat{s}_\alpha) k}\right) - \sum_{\gamma}{\epsilon_{\alpha\delta\gamma}d_{jk}^{J\gamma} }\nonumber\\
    &=\frac{i}{\hbar}\mel{j}{[\hat{l}_\alpha+\hat{s}_\alpha,\nabla_{J\delta}]}{k} - \sum_{\gamma}{\epsilon_{\alpha\delta\gamma}d_{jk}^{J\gamma}}
\end{align}
As above, since $\hat{l}_\alpha+\hat{s}_\alpha$ is an electronic operator that must commute with $\nabla_{J\delta}$, the first term is zero. In the second term, by replacing the dummy index $\gamma$ by $\zeta$, we arrive at
\begin{align}
    \sum_{I,\beta,\gamma}{\epsilon_{\alpha\beta\gamma}\nabla_{I\gamma}d^{J\delta}_{jk}} + \sum_{\zeta}{\epsilon_{\alpha\delta\zeta}d_{jk}^{J\zeta}} = 0
\end{align}
This concludes the proof.

\section{Inclusion of the Second Order Derivative Coupling Terms} \label{sec:second_order}

Below we will show that when Eq.~\eqref{eq:HnaGamma} is replaced by
\begin{align}
    H_{jk} = V_{jk} - i\hbar\sum_{I,\alpha}{\frac{P_{I\alpha}}{M_I}{\Gamma^{I\alpha}_{jk}}} - \hbar^2\sum_{I,\alpha,l}{\frac{\Gamma^{I\alpha}_{jl}\Gamma^{I\alpha}_{lk}}{2M_I}} \label{eq:HnaGamma2},
\end{align}
the resulting PSSH equations of motion still conserves the molecular linear and angular momentum during propagation.

Below we will assume the trajectory is propagating on the phase-space adiabat $\tilde{n}$. According to the Hamilton's equation, Eq.~\eqref{eq:psshdmR} remains unchanged, but Eq.~\eqref{eq:psshdmP} now becomes
\begin{align}
\label{eq:app:joe}
    \dot{P}_{I\alpha} &= -\tr[\sigma^{[\tilde{n}]}\nabla_{I\alpha} H] \nonumber\\
    &= -\tr[\sigma^{[\tilde{n}]}\left(\nabla_{I\alpha} V - i\hbar\sum_{J,\delta}{\frac{P_{J\delta}}{M_J}\nabla_{I\alpha}\Gamma_{J\delta}} - \hbar^2\sum_{J,\delta}{\frac{[\Gamma_{J\delta},\nabla_{I\alpha}\Gamma_{J\delta}]_+}{2M_J}}\right)]
\end{align}
where $[\cdot,\cdot]_+$ stands for the matrix anticommutator.

As compared against Eq.~\eqref{eq:psshdmP}, the only difference in Eq. \ref{eq:app:joe} is the extra anticommutator term, and so it makes sense to follow the derivations above in Sec. \ref{sec:cons:big} above.

\begin{itemize}
    \item For the case of linear momentum conservation, we follow the derivation in Sec.~\ref{sec:pssh_ptot} (using the expression in Eq.~\eqref{eq:dPmol_pssh2}), and when Hamiltonian \eqref{eq:HnaGamma2} is used instead of the Hamiltonian in Eq. \eqref{eq:HnaGamma}, we find
\begin{align}
    \dv{P_{mol,\alpha}}{t} =-\sum_I{\tr[\sigma^{[\tilde{n}]}\left(\nabla_{I\alpha} V - i\hbar\sum_{J,\delta}{\frac{P_{J\delta}}{M_J}\nabla_{I\alpha} \Gamma_{J\delta}} - \hbar^2\sum_{J,\delta}{\frac{[\Gamma_{J\delta},\nabla_{I\alpha}\Gamma_{J\delta}]_+}{2M_J}}\right)]} 
    \label{eq:p_app_almost}
\end{align}
The first two terms in Eq. \ref{eq:p_app_almost} are discussed in Sec.~\ref{sec:pssh_ptot}. The last term is zero since $\sum_I{\nabla_{I\alpha}\Gamma_{J\delta}} = 0$. Therefore the total molecular momentum is conserved.

\item For the case of angular momentum, we follow the derivation in Sec.~\ref{sec:pssh_ltot}
(using the expression in Eq.~\eqref{eq:dJmol_pssh2.5}),
and 
when Hamiltonian \eqref{eq:HnaGamma2} is used instead of the Hamiltonian in Eq. \eqref{eq:HnaGamma},
we find
\begin{align}
\label{eq:l_app_almost}
    \dv{L_{mol,\alpha}}{t} 
    &=\sum_{I,\beta,\gamma}{\epsilon_{\alpha\beta\gamma}\tr\left[\sigma^{[\tilde{n}]}\left(-i\hbar\Gamma_{I\beta}\frac{P_{I\gamma}}{M_I} \right.\right.} \nonumber\\
    &\qquad\left.\left.+i\hbar R_{I\beta}\sum_{J,\delta}{\frac{P_{J\delta}}{M_J}\nabla_{I\gamma}\Gamma_{J\delta}}+\hbar^2 R_{I\beta}\sum_{J,\delta}{\frac{[\Gamma_{J\delta},\nabla_{I\gamma}\Gamma_{J\delta}]_+}{2M_J}}\right)\right]
\end{align}
As discussed in Sec.~\ref{sec:pssh_ltot}, the first two terms on the RHS of Eq. \ref{eq:l_app_almost} evaluate to zero. Therefore we are left with
\begin{align}
    \dv{L_{mol,\alpha}}{t} &=\hbar^2\sum_{I,J,\beta,\gamma,\delta}{\epsilon_{\alpha\beta\gamma}R_{I\beta}\tr[\sigma^{[\tilde{n}]}\frac{[\Gamma_{J\delta},\nabla_{I\gamma}\Gamma_{J\delta}]_+}{2M_J}]} 
\end{align}
By Eq.~\eqref{eq:GammagradRcov} we have
\begin{align}
    \dv{L_{mol,\alpha}}{t} &=-\hbar^2\sum_{J,\delta}{\epsilon_{\alpha\delta\zeta}\tr[\sigma^{[\tilde{n}]}\frac{[\Gamma_{J\delta},\Gamma_{J\zeta}]_+}{2M_J}]} 
\end{align}
Because the expression in the trace above is symmetric between $\delta$ and $\zeta$, in the end the term is zero -- which indicates that running PSSH with Hamiltonian \eqref{eq:HnaGamma2} conserves the total molecular angular momentum.
\end{itemize}

\section{Proof of Eq.~\eqref{eq:bogradTinv} and \eqref{eq:bogradRinv}} \label{sec:gradinvproof}

In this section,  we will prove Eq.~\eqref{eq:bogradTinv} of the main text (and the proof of Eq.~\eqref{eq:bogradRinv} follows by an analogous procedure). If we expand $V_{jk} = \braket{j}{\hat{V}k}$ and apply the del operator to each term, we find
\begin{align}
    \sum_{I}{\nabla_{I\alpha}V_{jk}} = \sum_{I}{\braket{\nabla_{I\alpha}j}{\hat{V}k} + \bra{j}\nabla_{I\alpha}(\ket{\hat{V} k})}
\end{align}
By replacing the del operator by $\mathcal{P}$, and noting that $\sum_I{\bra{\nabla_{I\alpha}j}} = (\sum_I{\ket{\nabla_{I\alpha} j}})^\dagger = (\frac{i}{\hbar}\ket{\mathcal{P}_{\red{\alpha}}j})^\dagger = -\frac{i}{\hbar}\bra{\mathcal{P}_{\red{\alpha}}j}$, we find:
\begin{align}
    \sum_{I}{\nabla_{I\alpha}V_{jk}} &= -\frac{i}{\hbar}\mel{\mathcal{P}_\alpha j}{\hat{V}}{k} + \frac{i}{\hbar}\bra{j}\mathcal{P}_\alpha(\ket{\hat{V}k}) \nonumber\\
    &= -\frac{i}{\hbar}\mel{\mathcal{P}_\alpha j}{\hat{V}}{k} + \frac{i}{\hbar}\mel{j}{\hat{V}}{\mathcal{P}_\alpha k} + \frac{i}{\hbar}\mel{j}{[\mathcal{P}_\alpha,\hat{V}]}{k}
\end{align}
If we substitute Eq.~\eqref{eq:Tcov} for both $\ket{j}$ and $\ket{k}$, we find
\begin{align}
    \sum_{I}{\nabla_{I\alpha}V_{jk}} &= \frac{i}{\hbar}\mel{\hat{p}_\alpha j}{\hat{V}}{k} - \frac{i}{\hbar}\mel{j}{\hat{V}}{\hat{p}_\alpha k} +\frac{i}{\hbar} \mel{j}{[\mathcal{P}_\alpha, \hat{V}]}{k} \nonumber\\
    &= \frac{i}{\hbar}\mel{j}{[\hat{p}_\alpha,\hat{V}]}{k}+ \frac{i}{\hbar}\mel{j}{[\mathcal{P}_\alpha,\hat{V}]}{k} = \frac{i}{\hbar}\mel{j}{[\hat{p}_\alpha + \mathcal{P}_\alpha,\hat{V}]}{k} \label{eq:grad_cov_proof}
\end{align}
Since our Hamiltonian is invariant to translation (Eq.~\eqref{eq:VTinv}), Eq.~\eqref{eq:grad_cov_proof} equals  zero.

%

\end{document}